\pdfoutput=1

\documentclass[12pt,a4paper]{article}

\usepackage{ifthen} 
\newboolean{pdflatex}
\setboolean{pdflatex}{true} 

\newboolean{articletitles}
\setboolean{articletitles}{true} 

\newboolean{uprightparticles}
\setboolean{uprightparticles}{false} 

\newboolean{inbibliography}
\setboolean{inbibliography}{false} 

\newboolean{wordcount}
\setboolean{wordcount}{false} 

\RequirePackage[normalem]{ulem} 
\RequirePackage{color}\definecolor{RED}{rgb}{1,0,0}\definecolor{BLUE}{rgb}{0,0,1} 

\def\paperauthors{LHCb collaboration} 
\def\paperasciititle{First measurements of angular distribution of Z-> mu mu events in the forward region of pp collisions at the LHC} 
\def\papertitle{First measurement of the \ZmumuDecay angular coefficients in the forward region of $pp$ collisions at $\sqrt{s}=13\tev$}

\def\paperkeywords{{High Energy Physics}, {LHCb}} 
\def\papercopyright{\the\year\ CERN for the benefit of the LHCb collaboration} 
\def\paperlicence{CC BY 4.0 licence}
\def\paperlicenceurl{https://creativecommons.org/licenses/by/4.0/}

\usepackage[top=1in, bottom=1.25in, left=1in, right=1in]{geometry}

\usepackage{microtype}
\usepackage{lineno}  
\usepackage{xspace} 
\usepackage{caption} 

\usepackage{graphicx}  
\usepackage{color}
\usepackage{colortbl}
\graphicspath{{./figs/}} 

\usepackage{amsmath} 
\usepackage{amssymb}
\usepackage{amsfonts}
\usepackage{upgreek} 

\newcommand*\patchAmsMathEnvironmentForLineno[1]{%
\expandafter\let\csname old#1\expandafter\endcsname\csname #1\endcsname
\expandafter\let\csname oldend#1\expandafter\endcsname\csname
end#1\endcsname
 \renewenvironment{#1}%
   {\linenomath\csname old#1\endcsname}%
   {\csname oldend#1\endcsname\endlinenomath}%
}
\newcommand*\patchBothAmsMathEnvironmentsForLineno[1]{%
  \patchAmsMathEnvironmentForLineno{#1}%
  \patchAmsMathEnvironmentForLineno{#1*}%
}
\AtBeginDocument{%
\patchBothAmsMathEnvironmentsForLineno{equation}%
\patchBothAmsMathEnvironmentsForLineno{align}%
\patchBothAmsMathEnvironmentsForLineno{flalign}%
\patchBothAmsMathEnvironmentsForLineno{alignat}%
\patchBothAmsMathEnvironmentsForLineno{gather}%
\patchBothAmsMathEnvironmentsForLineno{multline}%
\patchBothAmsMathEnvironmentsForLineno{eqnarray}%
}

\usepackage{hyperxmp}

\usepackage[pdftex,
            pdfauthor={\paperauthors},
            pdftitle={\paperasciititle},
            pdfkeywords={\paperkeywords},
            pdfcopyright={Copyright (C) \papercopyright},
            pdflicenseurl={\paperlicenceurl}]{hyperref}

\usepackage[colorinlistoftodos,textsize=scriptsize]{todonotes}

\usepackage[bottom,flushmargin,hang,multiple]{footmisc}

\usepackage[all]{hypcap} 

\usepackage{xspace} 
\usepackage{upgreek}

\def\lhcb   {\mbox{LHCb}\xspace}
\def\atlas  {\mbox{ATLAS}\xspace}
\def\cms    {\mbox{CMS}\xspace}

\def\babar  {\mbox{BaBar}\xspace}

\def\cdf    {\mbox{CDF}\xspace}

\def\MagUp {\mbox{\em Mag\kern -0.05em Up}\xspace}

\ifthenelse{\boolean{uprightparticles}}%
{

 \def\Pmu         {\ensuremath{\upmu}\xspace}

 \def\PDelta      {\ensuremath{\Delta}\xspace}                 
 \def\PXi         {\ensuremath{\Xi}\xspace}                 
 \def\PLambda     {\ensuremath{\Lambda}\xspace}                 
 \def\PSigma      {\ensuremath{\Sigma}\xspace}                 
 \def\POmega      {\ensuremath{\Omega}\xspace}                 
 \def\PUpsilon    {\ensuremath{\Upsilon}\xspace}

 \def\PB      {\ensuremath{\mathrm{B}}\xspace}                 
                  
 \def\PD      {\ensuremath{\mathrm{D}}\xspace}

 \def\PK      {\ensuremath{\mathrm{K}}\xspace}

 \def\PW      {\ensuremath{\mathrm{W}}\xspace}

 \def\PZ      {\ensuremath{\mathrm{Z}}\xspace}                 
                  
 \def\Pb      {\ensuremath{\mathrm{b}}\xspace}                 
 \def\Pc      {\ensuremath{\mathrm{c}}\xspace}

 \def\Pi      {\ensuremath{\mathrm{i}}\xspace}

 \def\Ps      {\ensuremath{\mathrm{s}}\xspace}

 \def\thebaroffset{0.0em}
}
{

 \def\Pmu         {\ensuremath{\mu}\xspace}

 \mathchardef\PDelta="7101
 \mathchardef\PXi="7104
 \mathchardef\PLambda="7103
 \mathchardef\PSigma="7106
 \mathchardef\POmega="710A
 \mathchardef\PUpsilon="7107
                  
 \def\PB      {\ensuremath{B}\xspace}                 
                  
 \def\PD      {\ensuremath{D}\xspace}

 \def\PK      {\ensuremath{K}\xspace}

 \def\PW      {\ensuremath{W}\xspace}

 \def\PZ      {\ensuremath{Z}\xspace}                 
                  
 \def\Pb      {\ensuremath{b}\xspace}                 
 \def\Pc      {\ensuremath{c}\xspace}

 \def\Pi      {\ensuremath{i}\xspace}

 \def\Ps      {\ensuremath{s}\xspace}

 \def\thebaroffset{0.18em}
}
\newcommand{\offsetoverline}[2][\thebaroffset]{\kern #1\overline{\kern -#1 #2}}%

\makeatletter
\ifcase \@ptsize \relax
  \newcommand{\miniscule}{\@setfontsize\miniscule{4}{5}}
\or
  \newcommand{\miniscule}{\@setfontsize\miniscule{5}{6}}
\or
  \newcommand{\miniscule}{\@setfontsize\miniscule{5}{6}}
\fi
\makeatother

\DeclareRobustCommand{\optbar}[1]{\shortstack{{\miniscule (\rule[.5ex]{1.25em}{.18mm})}
  \\ [-.7ex] $#1$}}

\def\mup        {{\ensuremath{\Pmu^+}}\xspace}
\def\mun        {{\ensuremath{\Pmu^-}}\xspace} 

\def\ellm       {{\ensuremath{\ell^-}}\xspace}
\def\ellp       {{\ensuremath{\ell^+}}\xspace}

\def\W      {{\ensuremath{\PW}}\xspace}

\def\Z      {{\ensuremath{\PZ}}\xspace}

\def\squark    {{\ensuremath{\Ps}}\xspace}

\def\cquark    {{\ensuremath{\Pc}}\xspace}

\def\bquark    {{\ensuremath{\Pb}}\xspace}

\def\KorKbar {\kern \thebaroffset\optbar{\kern -\thebaroffset \PK}{}\xspace}

\def\D       {{\ensuremath{\PD}}\xspace}

\def\DorDbar {\kern \thebaroffset\optbar{\kern -\thebaroffset \PD}\xspace}

\def\Dp      {{\ensuremath{\D^+}}\xspace}
\def\Dm      {{\ensuremath{\D^-}}\xspace}

\def\DpDm    {\ensuremath{\Dp {\kern -0.16em \Dm}}\xspace}

\def\B       {{\ensuremath{\PB}}\xspace}

\def\BorBbar {\kern \thebaroffset\optbar{\kern -\thebaroffset \PB}\xspace}

\def\Bd      {{\ensuremath{\B^0}}\xspace}

\def\BdorBdbar {\kern \thebaroffset\optbar{\kern -\thebaroffset \Bd}\xspace}

\def\Bs      {{\ensuremath{\B^0_\squark}}\xspace}

\def\BsorBsbar {\kern \thebaroffset\optbar{\kern -\thebaroffset \Bs}\xspace}

\def\Y#1S{\ensuremath{\PUpsilon{(#1S)}}\xspace}

\def\LorLbar     {\kern \thebaroffset\optbar{\kern -\thebaroffset \PLambda}\xspace}

\def\to                 {\ensuremath{\rightarrow}\xspace}

\def\AT#1     {\ensuremath{A_{\mathrm{T}}^{#1}}\xspace}           

\def\C#1      {\ensuremath{\mathcal{C}_{#1}}\xspace}                       
\def\Cp#1     {\ensuremath{\mathcal{C}_{#1}^{'}}\xspace}                    
\def\Ceff#1   {\ensuremath{\mathcal{C}_{#1}^{\mathrm{(eff)}}}\xspace}        
\def\Cpeff#1  {\ensuremath{\mathcal{C}_{#1}^{'\mathrm{(eff)}}}\xspace}       
\def\Ope#1    {\ensuremath{\mathcal{O}_{#1}}\xspace}                       
\def\Opep#1   {\ensuremath{\mathcal{O}_{#1}^{'}}\xspace}                    

\newcommand{\aunit}[1]{\ensuremath{\text{\,#1}}}       

\newcommand{\tev}{\aunit{Te\kern -0.1em V}\xspace}
\newcommand{\gev}{\aunit{Ge\kern -0.1em V}\xspace}
\newcommand{\mev}{\aunit{Me\kern -0.1em V}\xspace}
\newcommand{\kev}{\aunit{ke\kern -0.1em V}\xspace}
\newcommand{\ev}{\aunit{e\kern -0.1em V}\xspace}
 
\newcommand{\mevc}{\ensuremath{\aunit{Me\kern -0.1em V\!/}c}\xspace}
\newcommand{\gevc}{\ensuremath{\aunit{Ge\kern -0.1em V\!/}c}\xspace}
\newcommand{\mevcc}{\ensuremath{\aunit{Me\kern -0.1em V\!/}c^2}\xspace}
\newcommand{\gevcc}{\ensuremath{\aunit{Ge\kern -0.1em V\!/}c^2}\xspace}

\def\fb   {\ensuremath{\aunit{fb}}\xspace}
\def\invfb   {\ensuremath{\fb^{-1}}\xspace}

\def\gsim{{~\raise.15em\hbox{$>$}\kern-.85em
          \lower.35em\hbox{$\sim$}~}\xspace}
\def\lsim{{~\raise.15em\hbox{$<$}\kern-.85em
          \lower.35em\hbox{$\sim$}~}\xspace}

\def\pt         {\ensuremath{p_{\mathrm{T}}}\xspace}

\def\geant      {\mbox{\textsc{Geant4}}\xspace}

\def\photos     {\mbox{\textsc{Photos}}\xspace}
\def\powheg     {\mbox{\textsc{Powheg}}\xspace}
\def\pythia     {\mbox{\textsc{Pythia8}}\xspace}
\def\resbos     {\mbox{\textsc{ResBos}}\xspace}

\def\tell1  {TELL1\xspace}
\def\ukl1   {UKL1\xspace}

\usepackage{cite} 
\usepackage{mciteplus}

\usepackage{longtable} 

\usepackage{xspace} 
\usepackage{upgreek}

\def\Cs  {\ensuremath{\frac{\rm{d}\sigma}{\rm{d}\cos\theta d\phi}}\xspace}
\def\Cscon  {\ensuremath{(1+\cos^{2}\theta)}\xspace}
\def\Cszero {\ensuremath{\frac{1}{2}A_{0}(1-3\cos^{2}\theta)}\xspace}
\def\Csone {\ensuremath{A_{1}\sin2\theta\cos\phi}\xspace}
\def\Cstwo {\ensuremath{\frac{1}{2}A_{2}\sin^{2}\theta\cos2\phi}\xspace}
\def\Csthree {\ensuremath{A_{3}\sin\theta\cos\phi}\xspace}
\def\Csfour {\ensuremath{A_{4}\cos\theta}\xspace}
\def\Csfive {\ensuremath{A_{5}\sin^{2}\theta\sin2\phi}\xspace}
\def\Cssix {\ensuremath{A_{6}\sin2\theta\sin\phi}\xspace}
\def\Csseven {\ensuremath{A_{7}\sin\theta\sin\phi}\xspace}

\def\ZllxFullDecay {\ensuremath{pp \to \gamma^{*}/Z +X \to \ellp \ellm + X}\xspace}

\def\ZmumuDecay {\ensuremath{Z \to \mup \mun}\xspace}

\newcommand{\TeVnosp}{\ifthenelse{\boolean{inbibliography}}{\ensuremath{~T\kern -0.05em eV}}{\ensuremath{\mathrm{Te\kern -0.1em V}}}}
\newcommand{\GeVnosp}{\ensuremath{\mathrm{Ge\kern -0.1em V}}}
\newcommand{\MeVnosp}{\ensuremath{\mathrm{Me\kern -0.1em V}}}
\newcommand{\keVnosp}{\ensuremath{\mathrm{ke\kern -0.1em V}}}
\newcommand{\eVnosp}{\ensuremath{\mathrm{e\kern -0.1em V}}}
\newcommand{\GeVcnosp}{\ensuremath{{\mathrm{Ge\kern -0.1em V\!/}c}}}
\newcommand{\MeVcnosp}{\ensuremath{{\mathrm{Me\kern -0.1em V\!/}c}}}
\newcommand{\GeVccnosp}{\ensuremath{{\mathrm{Ge\kern -0.1em V\!/}c^2}}}
\newcommand{\MeVccnosp}{\ensuremath{{\mathrm{Me\kern -0.1em V\!/}c^2}}}
\newcommand{\GeVGeVccccnosp}{\ensuremath{{\mathrm{Ge\kern -0.1em V^2\!/}c^4}}}

\newcommand{\lumiruntwo}{\nobreak{\ensuremath{ 5.1}}}
\usepackage{multirow}

\ifthenelse{\boolean{wordcount}}%

\begin{document}

\renewcommand{\thefootnote}{\fnsymbol{footnote}}
\setcounter{footnote}{1}


\begin{titlepage}
\pagenumbering{roman}

\vspace*{-1.5cm}
\centerline{\large EUROPEAN ORGANIZATION FOR NUCLEAR RESEARCH (CERN)}
\vspace*{1.5cm}
\noindent
\begin{tabular*}{\linewidth}{lc@{\extracolsep{\fill}}r@{\extracolsep{0pt}}}
\ifthenelse{\boolean{pdflatex}}
{\vspace*{-1.5cm}\mbox{\!\!\!\includegraphics[width=.14\textwidth]{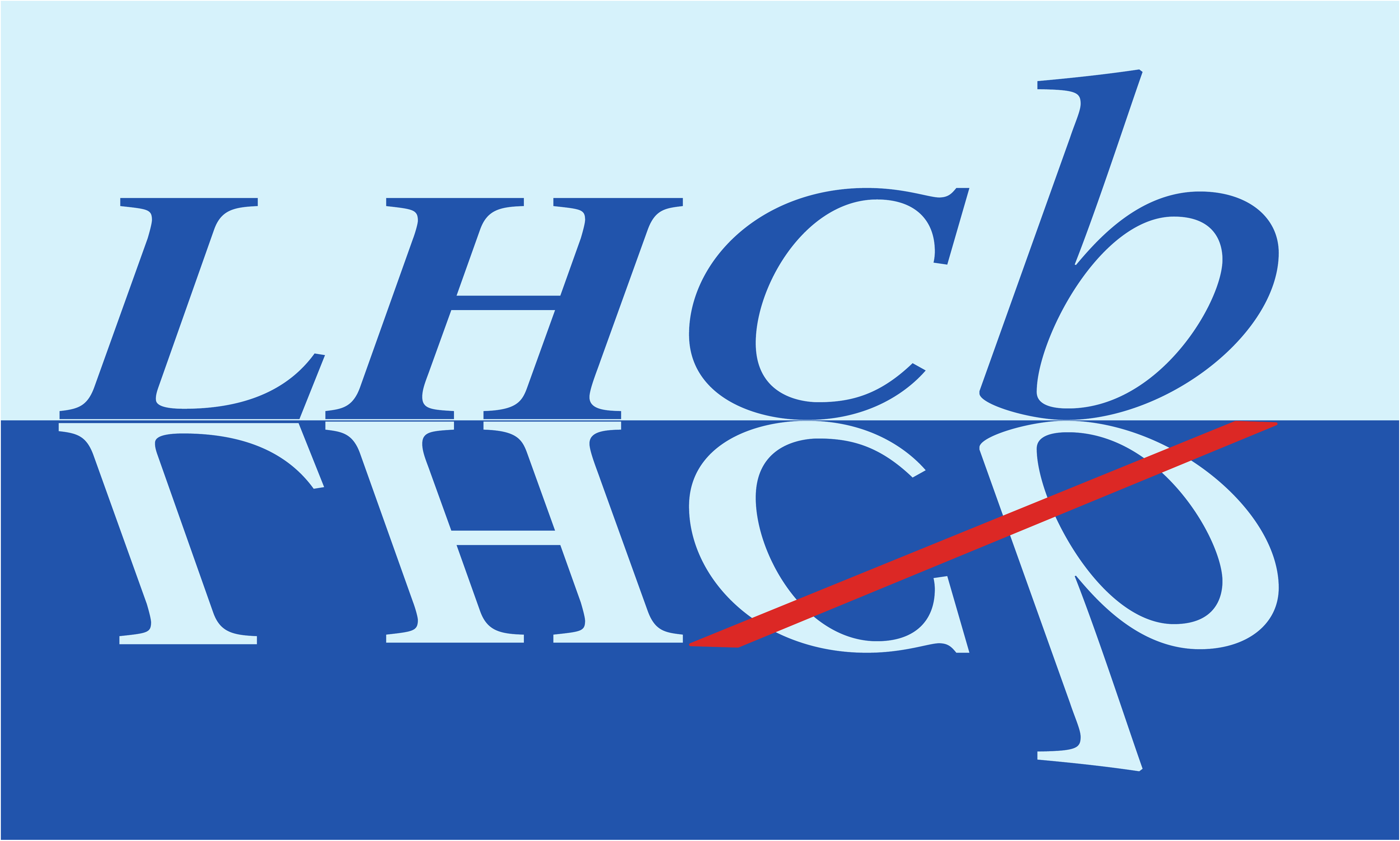}} & &}%
{\vspace*{-1.2cm}\mbox{\!\!\!\includegraphics[width=.12\textwidth]{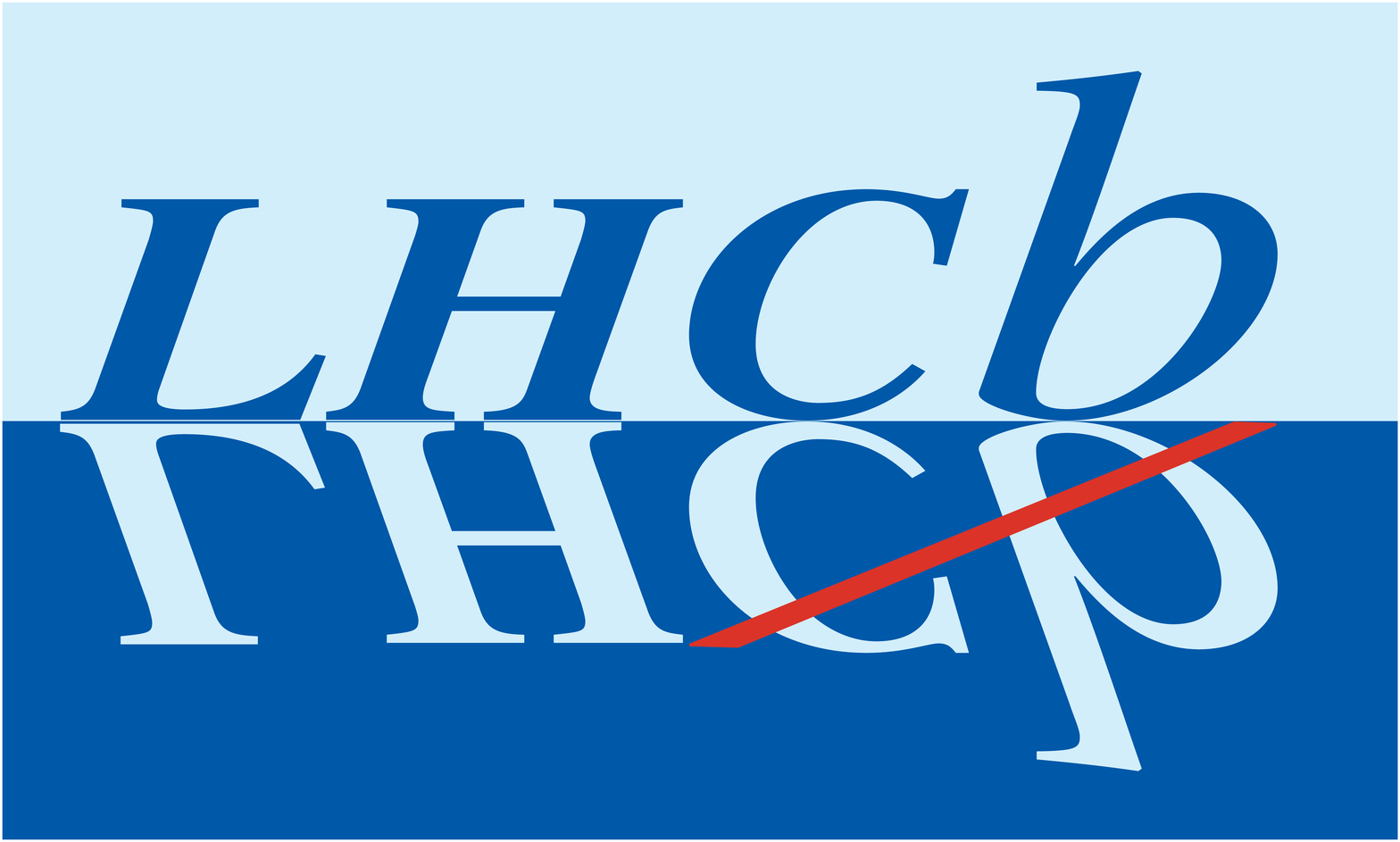}} & &}%
\\
 & & CERN-EP-2022-030 \\  
 & & LHCb-PAPER-2021-048 \\  
 & & \today \\ 
\end{tabular*}

\vspace*{4.0cm}

{\normalfont\bfseries\boldmath\huge
\begin{center}
  \papertitle 
\end{center}
}

\vspace*{2.0cm}

\begin{center}
\paperauthors\footnote{Authors are listed at the end of this Letter.}
\end{center}

\vspace{\fill}

\begin{abstract}
  \noindent
  The first study of the angular distribution of \mup\mun pairs produced in the forward rapidity region 
   via the Drell-Yan reaction \ZllxFullDecay is presented, 
   using data collected with the \lhcb detector at a center-of-mass energy of 13\tev, corresponding to an integrated luminosity of \lumiruntwo\invfb. 
   The coefficients of the five leading terms in the angular distribution 
   are determined as a function of the dimuon transverse momentum and rapidity. The results are compared to various theoretical predictions 
   of the \Z-boson production mechanism and can also be used to probe transverse-momentum-dependent parton distributions within the proton.
  
\end{abstract}

\vspace*{2.0cm}

\begin{center}
  Published in Phys. Rev. Lett. 129 (2022) 091801
\end{center}

\vspace{\fill}

{\footnotesize 
\centerline{\copyright~\papercopyright. \href{\paperlicenceurl}{\paperlicence}.}}
\vspace*{2mm}

\end{titlepage}


\newpage
\setcounter{page}{2}
\mbox{~}
%
%
%
%

\renewcommand{\thefootnote}{\arabic{footnote}}
\setcounter{footnote}{0}



\pagestyle{plain} 
\setcounter{page}{1}
\pagenumbering{arabic}


\noindent
Drell-Yan muon pairs ($\mu^+\mu^-$) produced near the \Z-boson mass pole provide an excellent source of information about electroweak (EW) parameters, 
probe the proton structure and test the underlying strong-interaction dynamics of \Z-boson production in proton-proton ($pp$) collisions, 
as described by quantum chromodynamics (QCD). 
With a dedicated detector instrumented in the forward region, the \lhcb experiment plays a unique role in the study of \ZmumuDecay processes 
in the forward rapidity ($y^Z$) region~\cite{LHCb-paper-2021-037}, 
especially for $y^Z>3.5$. Therefore, the \lhcb experiment is
complementary to the \atlas and \cms experiments, which are more efficient in the central region. 
The distributions of the \Z-boson transverse momentum, \pt, and the scattering angle, $\phi^*_{\eta}$, 
of the muons with respect to the proton beam direction in the rest frame of the dimuon system~\cite{phi_star} have already been measured by the \lhcb collaboration~\cite{LHCb-paper-2015-001,LHCb-paper-2015-049,LHCb-paper-2016-021,LHCb-paper-2021-037}. 
Key information is also encoded in the \ZmumuDecay angular distribution
in the forward region, which has not been fully explored.

In the Drell-Yan neutral-current process, 
the lepton pair is produced via a spin--1 gauge boson. 
The kinematic distribution of the final-state leptons ($\ell^{\pm}$) provides a direct probe of the polarization of the intermediate gauge boson,
which is in turn sensitive to the underlying QCD production mechanisms. The QCD mechanisms of the process \ZllxFullDecay, where $X$ represents any other particles, can be expressed using eight frame-dependent 
angular coefficients, $A_{i}$ ($i=0,...,7$). 
The coefficients depend on the invariant mass, \pt and rapidity of the lepton pair.
At Born level, the angular distribution of the leptons in the boson rest frame is given by~\cite{Mirkes:1990vn,Korner:1990im}
\begin{equation}
\label{eq:differentail}
\begin{aligned}
\Cs &\propto \Cscon + \Cszero +\Csone + \Cstwo\\
+&\Csthree + \Csfour +\Csfive + \Cssix +\Csseven,
\end{aligned}
\end{equation}
where $\theta$ and $\phi$ are the polar and azimuthal angles of the $\mu^+$ lepton in the Collins-Soper frame~\cite{CollinsSoperFrame1977}. At leading-order (LO) approximation within the framework of QCD, all angular coefficients vanish as the dilepton transverse momentum 
approaches zero, with the exception of $A_4$, which is nonzero due to parity violation in the weak interaction.
At next-to-leading-order (NLO), $A_{0-3}$ become nonzero, while $A_{5-7}$ have small deviations from zero 
only at next-to-next-to-leading-order (NNLO).

The equality $A_{0} = A_{2}$ is known as the Lam-Tung relation~\cite{Function:228}, 
which is valid for both $q\bar{q}$ and $gq$ processes at LO~\cite{Boer:2006eq}. However, small violations of the Lam-Tung relation occur in higher-order perturbative QCD calculations~\cite{Mirkes:1994eb,Gauld:2017tww},
as well as in QCD resummation calculations to all orders~\cite{Berger:2007si}. Therefore, precision measurements of $A_{0} -A_{2}$ can test the Lam-Tung relation. Nonperturbative effects can lead to violation of the Lam-Tung relation via a correlation between 
the transverse spin and transverse momentum of the initial-state quark or antiquark, 
represented by the Boer-Mulders transverse-momentum-dependent (TMD) parton distribution function (PDF)~\cite{Boer:1997nt,Boer:1999mm}. 
Significant violation of the Lam-Tung relation, with large $\cos{2\phi}$ modulations, has been observed in pion-induced Drell-Yan measurements~\cite{NA10:1986fgk,NA10:1987sqk,Conway:1989fs}. 
The $\cos 2\phi$ coefficient ($A_2$) in Eq.~(\ref{eq:differentail}) is proportional to the convolution of two Boer-Mulders functions, of the quark and the antiquark~\cite{Boer:1999mm}. 
Therefore, the angular measurements of the Drell-Yan process can improve constraints on nonperturbative partonic spin-momentum 
correlations within unpolarised protons via phenomenological extractions of the Boer-Mulders TMD PDF from these data in conjunction with data from other experiments.
The $Z$-boson cross-section measurements at low $p_T$ 
($\lesssim 20 \gevc$) by the \lhcb collaboration have already been used to extract the unpolarised TMD PDF~\cite{Bacchetta:2019sam}.

Previously, the \cdf collaboration published the first measurement~\cite{Aaltonen:2011nr} of 
the angular coefficients of lepton pairs produced near the \Z-boson mass pole at a hadron collider.
Measurements of the \W-boson angular coefficients have been published by the \cdf~\cite{PhysRevD.73.052002}, \cms~\cite{Chatrchyan:2011ig} and \atlas~\cite{ATLAS:2012au} collaborations.
The \cms collaboration measured $A_0$, $A_1$, $A_2$ and $A_4$ with \ZmumuDecay decays~\cite{Khachatryan:2015paa}, while the \atlas collaboration reported measurements of the complete set of coefficients~\cite{Aad:2016izn}.

This Letter reports the first measurement of the \ZmumuDecay angular coefficients in the forward rapidity region ($2 <y^Z < 5$) by the \lhcb experiment,
using \Z-boson candidates selected in the mass region \mbox{$50<M_{\mu\mu}<120\gevcc$}, 
where $M_{\mu\mu}$ is the invariant mass of the $\mu^+\mu^-$ pair. 
The parameters $A_{0}$, $A_{1}$, $A_{2}$ and $A_{3}$ are determined as a function of the \Z-boson \pt and $y^Z$, with the candidates selected in the mass region \mbox{$75<M_{\mu\mu}<105\gevcc$}. 
Events in the low mass range, \mbox{$50<M_{\mu\mu}<75\gevcc$}, and high mass range, \mbox{$105<M_{\mu\mu}<120\gevcc$}, are used to probe the Boer-Mulders function; 
these regions are dominated by contributions from virtual photons and their interference with the \Z-boson amplitude, 
with measurements in multiple mass regions adding sensitivity 
to the evolution of the TMD PDF with the hard scale.
With the higher-order QCD effects related to the \Z-boson \pt being integrated out,
the measurements of the angular coefficients 
as a function of $y^Z$ can be used to test the resummation calculations in the forward region.
Due to the relatively small data sample containing high-\pt ($>50\gevc$) \ZmumuDecay events, 
the coefficients $A_5$ to $A_7$ are fixed to zero. Since this study focuses on probing the QCD dynamics of the \Z-boson production, the $A_{4}$ coefficient, sensitive to the weak mixing angle~\cite{2006257}, is not reported. Nevertheless, in order to investigate its variation across the kinematic range, the difference with respect to its mean value, $\Delta A_{4}$, is measured. 

The \lhcb detector~\cite{Alves:2008zz,LHCb-DP-2014-002} is a
single-arm forward spectrometer covering the pseudorapidity range $2 <\eta < 5$, designed for
the study of particles containing \bquark\ or \cquark\ quarks. A silicon-strip vertex
detector~\cite{LHCb-DP-2014-001} surrounding the $pp$ interaction
region allows \cquark\ and \bquark\ hadrons to be identified from
their relatively long flight distance. 
A tracking system~\cite{LHCb-DP-2013-003} provides a measurement of momentum, $p$, 
of charged particles, and two imaging Cherenkov detectors~\cite{LHCb-DP-2012-003} 
are able to discriminate between different species of charged hadrons.
Photons, electrons and hadrons are identified by a calorimeter system consisting of
scintillating-pad and preshower detectors, an electromagnetic
and a hadronic calorimeter. 
Muons are identified by a system composed of alternating layers of iron and multiwire
proportional chambers~\cite{LHCb-DP-2012-002}.
The online event selection is performed by a trigger system~\cite{LHCb-DP-2012-004}, 
which consists of a hardware stage, based on information from the calorimeter and muon
systems~\cite{LHCb-DP-2020-001, LHCb-DP-2013-001}, 
followed by a software stage, which applies a full event
reconstruction incorporating near-real-time alignment and calibration of the detector~\cite{LHCb-PROC-2015-011}.

Simulated $pp$ collisions are generated using \pythia~\cite{Sjostrand:2007gs} with a specific \lhcb 
configuration~\cite{LHCb-PROC-2010-056}. The interaction of the generated particles with the detector, and its response,
are implemented using the \geant toolkit~\cite{Allison:2006ve} as described in Ref.~\cite{LHCb-PROC-2011-006}. 
Various theoretical predictions at the Born-level are compared with the measured results, 
including analytic resummed calculations, \resbos~\cite{Balazs:1997xd} and {\textsc{DYTurbo}}~\cite{Camarda:2019zyx}, 
and a fixed-order calculation with matching algorithms to veto the double counting of quantum electrodynamic (QED) final-state radiation (FSR) and parton shower, \powheg-BOX~\cite{Powheg:040,Powheg:2092,Powheg:4802,Powheg:043}. 
Predictions from the \resbos~\cite{Balazs:1997xd} generator are produced using the PDF known as CT14HERA2~\cite{CT142017}.
The \resbos calculation combines NLO fixed-order perturbative calculations at high boson \pt, with the soft-gluon
Collins-Soper-Sterman resummation formalism~\cite{Collins:1984kg,Collins:1981uk,Collins:1981va} 
at low boson \pt, 
with consideration of a nonperturbative contribution from the parton intrinsic transverse momentum, 
which is an all-orders summation of large terms from gluon emission.
The {\textsc{DYTurbo}}~\cite{Camarda:2019zyx} program is an NNLO generator for the calculation of the QCD
transverse-momentum resummation of Drell–Yan cross sections up to next-to-next-to-leading logarithmic accuracy combined with the 
fixed-order results. {\textsc{DYTurbo}} includes the full kinematic dependence of the lepton pair with the corresponding spin correlations and finite-width effects.
\powheg-BOX is an NLO generator and can be interfaced with \pythia for QCD and EW showering, but without the angular damping factors.


Measurements are performed using 
\lumiruntwo\invfb of data at a center-of-mass energy of 13\tev collected by the \lhcb detector during the years 2016--2018.
For each \ZmumuDecay candidate, at least one of the muons is required to pass hardware and software single-muon trigger decision stages.
All muon candidates are required to have transverse momentum $\pt^{\mu}>20\gevc$ and be in the range $2.0<\eta<4.5$. 
The relative uncertainty in the momentum of the muon track must be less than $10\%$. 
To further suppress background sources, muon candidates are required to pass an isolation requirement 
which is parameterised as $z=\pt^{\mu}/\pt^{\mu-\rm{cone}}>0.85$, 
where $\pt^{\mu-\rm{cone}}$ is the sum of the muon \pt and that of the tracks within $\delta R<0.5$, 
and $\delta R$ is $\sqrt{\delta \eta^{2} + \delta \Phi^{2}}$. The quantities $\delta \eta$ and $\delta \Phi$ give the separation between the muon and neighbouring tracks 
in pseudorapidity $\eta$ and the laboratory azimuthal angle $\Phi$, respectively. 
Candidate \Z-bosons are composed of two muon candidates with opposite charge 
that originate from a common primary $pp$ interaction vertex. In total, 818\,074 (745\,343) \ZmumuDecay candidates are selected in the mass region $50<M_{\mu\mu}<120\gevcc$ ($75<M_{\mu\mu}<105\gevcc$).
Background contributions from heavy-flavour processes and misidentified hadrons are estimated using data control samples~\cite{LHCb-paper-2021-037}, 
while background sources from $W^{+}W^{-}$, $W^{\pm}Z$, $ZZ$, $Z\to\tau^{+}\tau^{-}$ and $t\bar{t}$ processes 
are estimated from simulation. The total background contribution is determined to be $0.2\%$.

Detector misalignment effects are studied and suppressed using the 
\ZmumuDecay data events, where the mass peak position of the selected
\Z-boson candidates is calibrated in different kinematic and geometric regions to the
world average value~\cite{PDG2020}.
To improve the agreement between data and simulation, 
selection efficiencies determined from simulation are corrected to 
corresponding values measured in the data. 
The selection efficiencies include the muon trigger, muon track reconstruction, muon isolation and identification.
These efficiencies are estimated using \ZmumuDecay data and simulation events, 
with the tag-and-probe method~\cite{LHCb-paper-2021-037}, as functions of muon
$\eta$ and $\Phi$. 
Additionally, corrections to the momentum scale and resolution are applied to the simulation.

The angular coefficients are determined in six intervals of the \Z-boson \pt and five intervals of $y^Z$, 
by fitting the two decay angles $\cos\theta$ and $\phi$ of the selected \ZmumuDecay candidates, 
using an unbinned maximum-likelihood method in each interval. 
The probability density function used in the fit consists of a
signal-only function, obtained from Eq.~(\ref{eq:differentail}),
convoluted with the detector resolution and acceptance effects. 
The angular coefficients ($A_{0-4}$) are the only free parameters in the fit,
while the background contributions are effectively subtracted from the data sample by including background events 
with negative weights derived from simulation and data control samples. 
Simulated \pythia events are used to study the background modeling along with the detector resolution and acceptance effects, 
with FSR turned on at the generator level, thereby including the Born/Bare corrections~\cite{Placzek:2003zg}. 
This way, the measured $A_{i}$ coefficients are corrected to the Born level. 
The normalization weights method is used to avoid computing-intensive numerical integration of the probability density function in the fit. 
This method was introduced and used by the \babar collaboration~\cite{Aubert:2001pe,Aubert:2004cp}, 
and has been extensively used in \lhcb heavy-flavour studies~\cite{LHCb-PAPER-2011-021}. 
To avoid potential bias from the input angular coefficients in the simulation, an iterative method is used.
The measured results are found to be stable after four iterations.

Several sources of systematic uncertainty are considered. Simulation is used to determine the detector acceptance effects in the fit process. 
To estimate the uncertainty from the finite number of simulated events, the bootstrap method~\cite{efron:1979} is used. This uncertainty varies from 0.0016 to 0.0402 in different kinematic regions, and is
the dominant systematic uncertainty. 
Another significant uncertainty arises from the selection efficiencies. 
In the determination of the uncertainty from the efficiency corrections, 
two independent sources are considered: an uncertainty from the size of the control sample and one from the kinematic dependence of the corrections. 
For the second uncertainty, the default two-dimensional (muon $\eta$ and $\phi$) efficiency corrections are replaced with
one-dimensional corrections, and variations of the results are taken as uncertainties. 
The fit method is validated with simulation samples. 
The angular coefficients of the simulation
are changed to values far from the default (increased
by a factor of 1.5, or set to 0).
In the validation, the new angular coefficients are compared with the values from the fit, 
and differences between them are taken as uncertainties. 
The uncertainty from the muon momentum scaling and smearing are calculated by varying the corrections by their uncertainties, and the changes in the measured results are taken as the associated systematic uncertainties. These are found to be negligible. 

The systematic uncertainties associated to the background are estimated by varying the background contribution by
$\pm \sigma$, where $\sigma$ includes the uncertainties from the theoretical predictions of 
the cross-sections and the statistical uncertainties from the simulation.
The systematic uncertainty from the FSR is estimated by comparing the
measured results using the simulation sample weighted to 
the \powheg prediction with different FSR algorithms of \photos and \pythia. 
Differences from different FSR approaches are taken as uncertainties. 
The systematic uncertainty from the PDF is estimated using 58 sub-PDF sets of CT18NNLO~\cite{Hou:2019efy}, 
with the prescription recommended by the PDF sets groups~\cite{Alekhin:2011sk}. 
All systematic uncertainties are considered as uncorrelated, and the total systematic uncertainty 
is taken as their sum in quadrature.

\begin{figure}[tb]
\begin{center}
\includegraphics[width=0.8\textwidth]{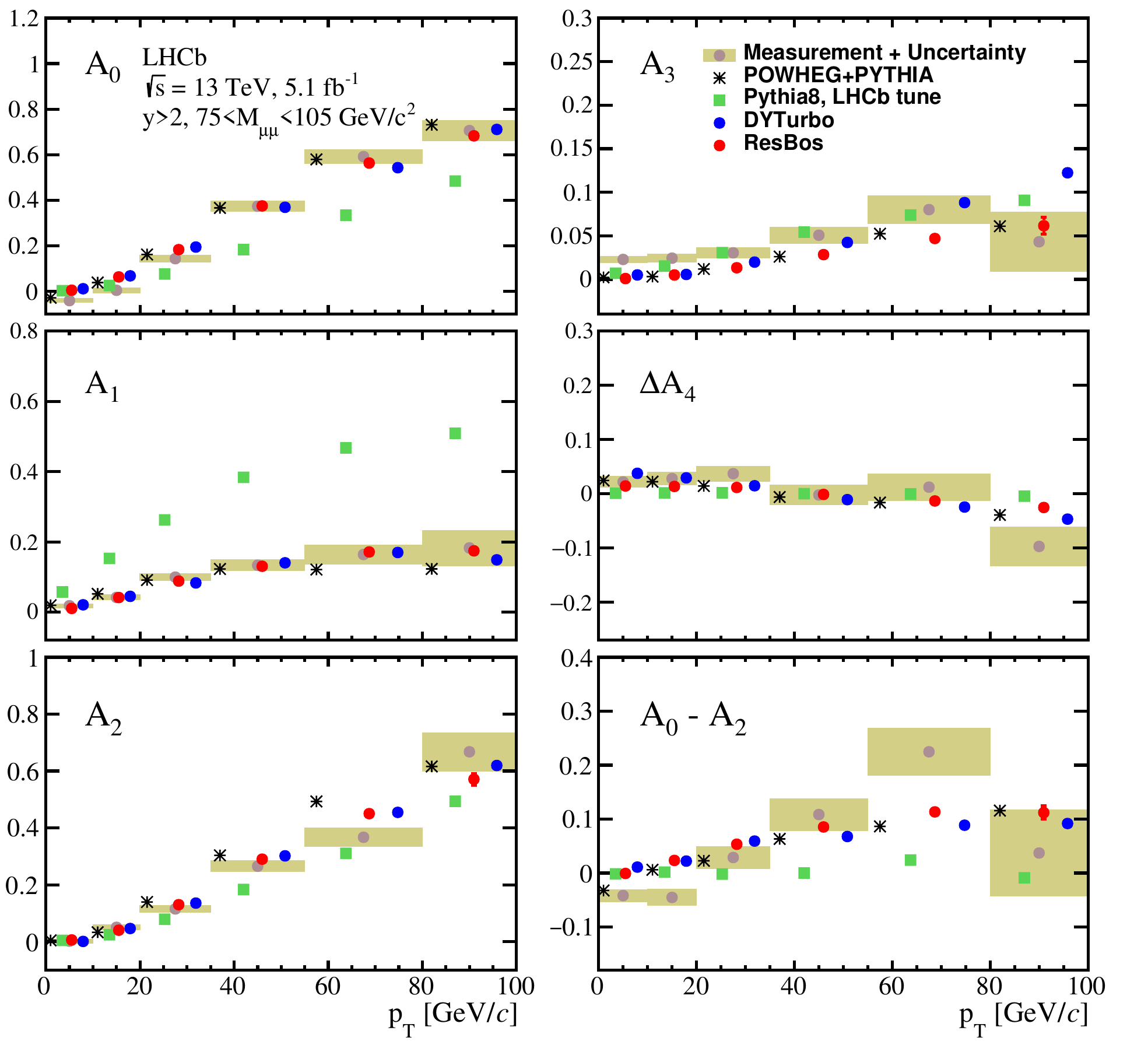}
\caption{Comparison of the measured angular coefficients with different predictions, as a function of the \Z-boson \pt, in the rapidity region of \Z-boson $y^Z>2$ and \mbox{$75<M_{\mu\mu}<105\gevcc$}. The total uncertainty (shown in the figure) is dominated by the statistical component. The theoretical predictions correspond to the same \Z-boson \pt bins as data and the \pt shifts of the theoretical markers in all plots are for visualization purposes only. {\textsc{DYTurbo}} and \resbos predictions include the theoretical uncertainties.}
\label{fig:combine_pt}
\end{center}
\end{figure}

The measured angular coefficients $A_{0-3}$ as well as $\Delta A_{4}$ and the difference $A_{0}-A_{2}$, 
are presented as a function of the \Z-boson \pt in Fig.~\ref{fig:combine_pt}. The measurements are compared with the predictions of \resbos, {\textsc{DYTurbo}}, \powheg-BOX+\pythia, and \pythia with \lhcb tuning.
The measurement of $A_0$ is in good agreement with {\textsc{DYTurbo}}, \resbos, and \powheg-BOX predictions, 
but disfavors the \pythia results in the high \pt region ($>20\gevc$). 
However, the \pythia calculations show similar increasing trends as the measurements in the high \pt region. 
The measurement of $A_1$ disagrees with the \pythia result, but is in good agreement with other calculations in all \pt regions. 
Within large statistical uncertainties, the measurements of $A_2$ are mostly in reasonable agreement with predictions, while for $A_3$ there is a noticeable underestimation of most predictions up to 55\gevcc. 
The measured $\Delta A_{4}$ is in good agreement with theoretical predictions in all \pt regions. Finally, for the $A_0-A_2$ measurement, 
the \lhcb results are compatible with zero only in the highest \pt interval, but there the uncertainty is relatively large. Also, most predictions are not compatible with zero in the high-\pt intervals. This violation of the Lam-Tung relations measured by the \lhcb collaboration is consistent with previous 
results from the \cms~\cite{Khachatryan:2015paa} and \atlas collaborations ~\cite{Aad:2016izn}. 
The measured $A_0-A_2$ results are significantly smaller than
the predictions at low \pt ($<20\gevc$) and slightly smaller at high \pt ($>80\gevc$), 
while larger than the predictions in other \pt regions. 
A summary of the full results, including measurements in bins of $y^Z$, 
is provided in the Supplemental Material~\cite{ref:SupplMat}. 
With the exception of $A_{0}$, the measured angular coefficients do not vary significantly as a function of $y^Z$, when the measurement is performed for the full \pt range. 
There is reasonable agreement between the measurements and \resbos calculations for $A_{1-3}$ and $\Delta A_{4}$, while for $A_{0}$ there is tension between the measurements and predictions, especially in the highest $y^Z$ region. The $p$-value between measurements and theoretical predictions is calculated to be 0.06.
The differences between $A_{0}$ and the \resbos predictions in the $y^Z$ distribution indicate that there is a $y^Z$-dependence in the QCD resummation or additional higher-order effects. 
More detailed studies on the predictions as a function of $y^Z$ are necessary.

\begin{figure}[!htbp]
 \begin{center}
 \includegraphics[width=0.8\textwidth]{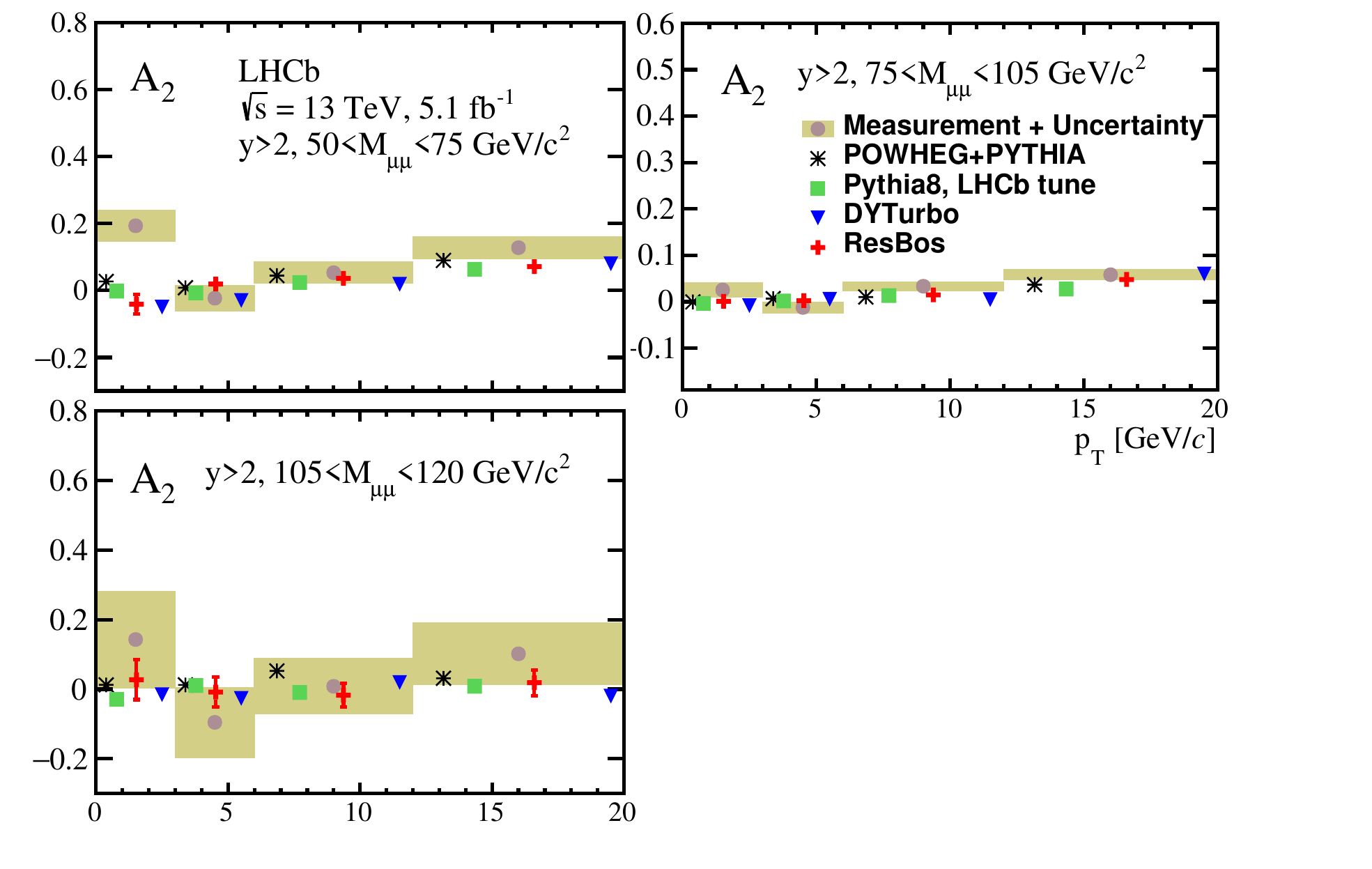}
\caption{Comparison of the measured angular coefficient $A_{2}$ with different predictions, as a function of the \Z-boson \pt, in the rapidity region of \Z-boson $y^Z>2$ and \mbox{$50<M_{\mu\mu}<75\gevcc$} (left), \mbox{$50<M_{\mu\mu}<75\gevcc$} (middle) and \mbox{$105<M_{\mu\mu}<120\gevcc$} (right). The total uncertainty (shown in the figure) is dominated by the statistical component. The horizontal positions of the theoretical predictions within the bins are adjusted to increase visibility. {\textsc{DYTurbo}} and \resbos predictions include the theoretical uncertainties.}
\label{fig:lowpt_towmassregion}
\end{center}
\end{figure}

The nonperturbative Boer-Mulders TMD PDF can be probed with the measurement of $A_2$ in the lower \pt region. Measurements of $A_2$ in the low, middle and high $M_{\mu\mu}$ regions are shown in Fig.~\ref{fig:lowpt_towmassregion}, where the dimuon \pt region is divided into four intervals from \mbox{0--20\gevc}. Despite the limited sample size with the finer \pt intervals, several observations can be made. 
In the low $M_{\mu\mu}$, the measured $A_2$ value in the lowest dimuon \pt region ($\pt<3\gevc$) deviates significantly from all predictions. 
It is unclear if nonperturbative spin-momentum correlations in the proton, described by the Boer-Mulders distribution,
could lead to such variations as no phenomenological calculations are available. 
None of the calculations used in the comparison include this type of nonperturbative effect. 
In other \pt regions, reasonable agreement
between measurements and predictions is seen.

In summary, the first measurements of the angular coefficients of Drell-Yan $\mu^+\mu^-$ pairs
in the forward rapidity region of $pp$ collisions are presented. These quantities provide more information on the \Z-boson production compared to other
traditional observables. 
The measurements, which are given as a function of the \Z-boson \pt and $y^Z$, in both the \Z peak and lower and higher $M_{\mu\mu}$ regions, are compared with various predictions. 
Some tension between the measurements and the theoretical predictions appears in the lower $M_{\mu\mu}$ and low-\pt region, and in the $y^Z$ dependence (See the Supplemental Material~\cite{ref:SupplMat} for further details). More studies of the theoretical models are needed to shed light on the apparent discrepancies. This analysis provides important and unique inputs for future phenomenological extractions of spin-momentum correlations in the proton in terms of the Boer-Mulders TMD PDF and its evolution with the hard scale.

\section*{Acknowledgements}
%
%
\noindent We express our gratitude to our colleagues in the CERN
accelerator departments for the excellent performance of the LHC. We
thank the technical and administrative staff at the LHCb
institutes.
We acknowledge support from CERN and from the national agencies:
CAPES, CNPq, FAPERJ and FINEP (Brazil); 
MOST and NSFC (China); 
CNRS/IN2P3 (France); 
BMBF, DFG and MPG (Germany); 
INFN (Italy); 
NWO (Netherlands); 
MNiSW and NCN (Poland); 
MEN/IFA (Romania); 
MSHE (Russia); 
MICINN (Spain); 
SNSF and SER (Switzerland); 
NASU (Ukraine); 
STFC (United Kingdom); 
DOE NP and NSF (USA).
We acknowledge the computing resources that are provided by CERN, IN2P3
(France), KIT and DESY (Germany), INFN (Italy), SURF (Netherlands),
PIC (Spain), GridPP (United Kingdom), RRCKI and Yandex
LLC (Russia), CSCS (Switzerland), IFIN-HH (Romania), CBPF (Brazil),
PL-GRID (Poland) and NERSC (USA).
We are indebted to the communities behind the multiple open-source
software packages on which we depend.
Individual groups or members have received support from
ARC and ARDC (Australia);
AvH Foundation (Germany);
EPLANET, Marie Sk\l{}odowska-Curie Actions and ERC (European Union);
A*MIDEX, ANR, IPhU and Labex P2IO, and R\'{e}gion Auvergne-Rh\^{o}ne-Alpes (France);
Key Research Program of Frontier Sciences of CAS, CAS PIFI, CAS CCEPP, 
Fundamental Research Funds for the Central Universities, 
and Sci. \& Tech. Program of Guangzhou (China);
RFBR, RSF and Yandex LLC (Russia);
GVA, XuntaGal and GENCAT (Spain);
the Leverhulme Trust, the Royal Society
 and UKRI (United Kingdom).

\addcontentsline{toc}{section}{References}
\setboolean{inbibliography}{true}

\bibliographystyle{LHCb}
\bibliography{main,LHCb-PAPER,LHCb-CONF,LHCb-DP,LHCb-TDR,z-refs}
 
\ifthenelse{\boolean{wordcount}}%
\newpage


\clearpage



\section*{Supplemental material}
\label{app:tabled_results}

\begin{figure}[tb]
  \begin{center}
      \includegraphics[width=0.8\textwidth]{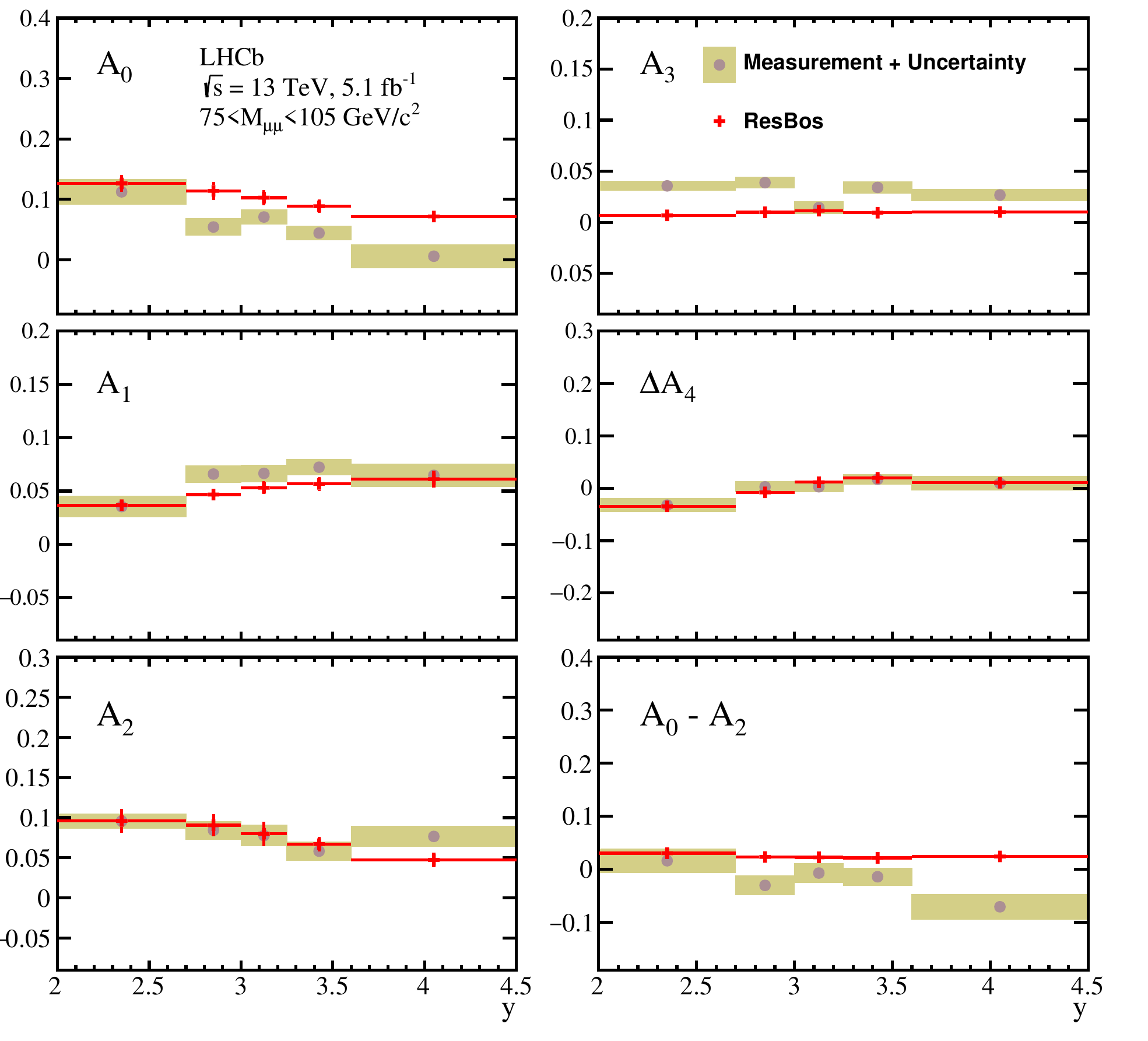}
\caption{Comparison of the measured angular coefficients 
with \resbos predictions as a function of the $y^Z$ in the \mbox{$75<M_{\mu\mu}<105\gevcc$} mass region. The total uncertainty (shown in the figure) is dominated by the statistical component. \resbos predictions include the theoretical uncertainties.}
  \label{fig:combine_y}
  \end{center}
\end{figure}

The number of data candidate events in different \Z-boson \pt and rapidity intervals are summarized in Table~\ref{tab:tab_events_y}. 
Summaries of measured angular coefficients 
$A_i$ are shown in Table~\ref{tab:sys-pt} in intervals of \Z-boson \pt, and in Table~\ref{tab:sys-y} and Fig.~\ref{fig:combine_y} in intervals of \Z-boson rapidity.
Summaries of measured angular coefficients and regularised uncertainties of $A_2$ in the \Z-boson low-\pt region, with events in different mass regions, 
are given in Table~\ref{tab:sys-a2}. 

\begin{table}[!htbp]
\caption{The number of selected data candidate events in different \Z boson \pt and rapidity intervals in the mass region $50<M_{\mu\mu}<120\gevcc$.}
\centering
\resizebox{!}{1.35cm}{
\begin{tabular}{cc|cc}
\hline
\pt interval & Yields & $y^Z$ interval & Yields \\
\hline
0-10 & 366305 & 2-2.7 & 203272\\
10-20& 214341 &2.7-3 & 171839\\
20-35 & 130525&3-3.25 & 152349\\
35-55 & 63285 &3.25-3.6 & 171900\\
55-80 & 27445&3.6-4.5 & 118714\\
80-100& 8120& &\\
\hline
\end{tabular}}
\label{tab:tab_events_y}
\end{table}

\begin{table}[!htbp]
\caption{Summary of measured angular coefficients and regularised uncertainties for $A_{i}$ and $A_{0}-A_2$, in intervals of \Z-boson \pt, in the region of \Z-boson $y>2$ and \mbox{$75<M_{\mu\mu}<105\gevcc$}. The total systematic uncertainty is shown with the breakdown into its underlying components. Entries marked with `-' indicate that the uncertainty is below 0.0001.}
\centering
\resizebox{!}{6.35cm}{
\begin{tabular}{c|c|c|c|c|c|c|c|c|c|c|c|c|}
\hline
\hline
 & \multicolumn{6}{c}{$\pt(Z) \in [0,10] [\gev]$} & \multicolumn{6}{|c|}{$\pt(Z) \in [10,20] [\gev]$}\\
\hline
\hline
Coefficient & $A_{0}$ & $A_{1}$ & $A_{2}$ & $A_{3}$ &$\Delta A_{4}$ & $A_{0}-A{2}$ & $A_{0}$ & $A_{1}$ & $A_{2}$ & $A_{3}$ &$\Delta A_{4}$ & $A_{0}-A{2}$\\
\hline
Total & -0.0401 & 0.0180 & 0.0019 & 0.0229 & 0.0219 & -0.0418 &0.0052 & 0.0417 & 0.0506 & 0.0244 & 0.0280 & -0.0452  \\
\hline
Stat & 0.0075 & 0.0046 & 0.0062 & 0.0029 & 0.0054 & 0.0097 &0.0101 & 0.0061 & 0.0083 & 0.0039 & 0.0073 & 0.0130  \\
\hline
Syst & 0.0040 & 0.0028 & 0.0036 & 0.0017 & 0.0030 & 0.0053 &0.0055 & 0.0037 & 0.0045 & 0.0022 & 0.0042 & 0.0071  \\
\hline
MC Stat & 0.0039 & 0.0025 & 0.0033 & 0.0016 & 0.0030 & 0.0051 &0.0054 & 0.0035 & 0.0044 & 0.0022 & 0.0041 & 0.0070  \\
FSR & - & - & - & - & - & - &- & - & - & - & - & -  \\
Eff & - & - & - & - & - & - &- & - & - & - & - & -  \\
Bkg & - & - & - & - & - & - &- & - & - & - & - & -  \\
Smear & - & - & - & - & - & - &- & - & - & - & - & -  \\
PDF & 0.0003 & 0.0003 & 0.0012 & 0.0002 & 0.0001 & 0.0010 &0.0003 & 0.0004 & 0.0007 & 0.0002 & 0.0006 & 0.0006  \\
Extraction & 0.0008 & 0.0010 & 0.0004 & - & - & 0.0008 &0.0007 & 0.0013 & 0.0004 & - & 0.0002 & 0.0008  \\
\hline
\hline
 & \multicolumn{6}{c}{$\pt(Z) \in [20,35] [\gev]$} & \multicolumn{6}{|c|}{$\pt(Z) \in [35,55] [\gev]$}\\
\hline
\hline
Coefficient & $A_{0}$ & $A_{1}$ & $A_{2}$ & $A_{3}$ &$\Delta A_{4}$ & $A_{0}-A{2}$ & $A_{0}$ & $A_{1}$ & $A_{2}$ & $A_{3}$ &$\Delta A_{4}$ & $A_{0}-A{2}$\\
\hline
Total & 0.1439 & 0.0995 & 0.1152 & 0.0305 & 0.0371 & 0.0288 &0.3740 & 0.1334 & 0.2653 & 0.0507 & -0.0021 & 0.1084  \\
\hline
Stat & 0.0138 & 0.0083 & 0.0110 & 0.0053 & 0.0096 & 0.0176 &0.0204 & 0.0134 & 0.0166 & 0.0082 & 0.0140 & 0.0263  \\
\hline
Syst & 0.0073 & 0.0044 & 0.0061 & 0.0028 & 0.0054 & 0.0095 &0.0105 & 0.0078 & 0.0090 & 0.0045 & 0.0081 & 0.0138  \\
\hline
MC Stat & 0.0071 & 0.0041 & 0.0060 & 0.0027 & 0.0054 & 0.0094 &0.0104 & 0.0075 & 0.0086 & 0.0044 & 0.0081 & 0.0134  \\
FSR & - & - & - & - & - & - &0.0003 & 0.0002 & 0.0002 & - & - & 0.0004  \\
Eff & 0.0001 & - & 0.0001 & - & - & 0.0001 &0.0001 & - & - & - & - & 0.0002  \\
Bkg & - & - & - & - & - & 0.0001 &- & - & - & - & - & -  \\
Smear & - & - & - & - & - & - &- & - & - & - & - & -  \\
PDF & 0.0009 & 0.0009 & 0.0007 & 0.0002 & 0.0003 & 0.0014 &0.0006 & 0.0004 & 0.0023 & 0.0009 & 0.0003 & 0.0019  \\
Extraction & 0.0010 & 0.0014 & 0.0006 & 0.0001 & 0.0001 & 0.0012 &0.0016 & 0.0021 & 0.0017 & 0.0005 & 0.0005 & 0.0023  \\
\hline
\hline
 & \multicolumn{6}{c}{$\pt(Z) \in [55,80] [\gev]$} & \multicolumn{6}{|c|}{$\pt(Z) \in [80,100] [\gev]$}\\
\hline
\hline
Coefficient & $A_{0}$ & $A_{1}$ & $A_{2}$ & $A_{3}$ &$\Delta A_{4}$ & $A_{0}-A{2}$ & $A_{0}$ & $A_{1}$ & $A_{2}$ & $A_{3}$ &$\Delta A_{4}$ & $A_{0}-A{2}$\\
\hline
Total & 0.5917 & 0.1639 & 0.3671 & 0.0801 & 0.0123 & 0.2247 &0.7061 & 0.1826 & 0.6676 & 0.0432 & -0.0971 & 0.0371  \\
\hline
Stat & 0.0261 & 0.0237 & 0.0275 & 0.0139 & 0.0200 & 0.0379 &0.0378 & 0.0442 & 0.0572 & 0.0293 & 0.0311 & 0.0686  \\
\hline
Syst & 0.0130 & 0.0132 & 0.0163 & 0.0079 & 0.0110 & 0.0209 &0.0217 & 0.0241 & 0.0351 & 0.0175 & 0.0167 & 0.0413  \\
\hline
MC Stat & 0.0121 & 0.0129 & 0.0160 & 0.0079 & 0.0110 & 0.0201 &0.0206 & 0.0233 & 0.0345 & 0.0174 & 0.0165 & 0.0402  \\
FSR & 0.0006 & 0.0001 & 0.0004 & - & 0.0001 & 0.0008 &0.0004 & 0.0008 & 0.0005 & 0.0002 & - & 0.0006  \\
Eff & 0.0006 & 0.0002 & 0.0003 & 0.0001 & 0.0002 & 0.0007 &0.0004 & 0.0004 & 0.0008 & 0.0005 & 0.0002 & 0.0009  \\
Bkg & 0.0002 & 0.0001 & 0.0002 & - & 0.0001 & 0.0003 &0.0005 & 0.0002 & 0.0008 & 0.0002 & 0.0002 & 0.0009  \\
Smear & - & - & - & - & - & - &- & - & - & - & - & -  \\
PDF & 0.0010 & 0.0017 & 0.0009 & 0.0006 & 0.0008 & 0.0016 &0.0011 & 0.0024 & 0.0041 & 0.0009 & 0.0015 & 0.0035  \\
Extraction & 0.0045 & 0.0021 & 0.0029 & 0.0008 & 0.0004 & 0.0054 &0.0067 & 0.0056 & 0.0054 & 0.0010 & 0.0019 & 0.0086  \\

\hline
\hline
\end{tabular}}
\label{tab:sys-pt}
\end{table}
\begin{table}[!htbp]
\caption{Summary of measured angular coefficients and regularised uncertainties for $A_{i}$ and $A_{0}-A_2$, in intervals of \Z-boson rapidity, in the region of \Z-boson \mbox{$\pt<100 \gevc$} and \mbox{$75<M_{\mu\mu}<105\gevcc$}. The total systematic uncertainty is
shown with the breakdown into its underlying components. Entries marked with `-' indicate that the uncertainty is
below 0.0001.}
\centering
\resizebox{!}{4.35cm}{
\begin{tabular}{c|c|c|c|c|c|c|c|c|c|c|c|c|}
\hline
\hline
 & \multicolumn{6}{c}{$y^Z \in [2,2.7]$} & \multicolumn{6}{|c|}{$y^Z \in [2.7,3]$}\\
\hline
\hline
Coefficient & $A_{0}$ & $A_{1}$ & $A_{2}$ & $A_{3}$ &$\Delta A_{4}$ & $A_{0}-A{2}$ & $A_{0}$ & $A_{1}$ & $A_{2}$ & $A_{3}$ &$\Delta A_{4}$ & $A_{0}-A{2}$\\
\hline
Total & 0.1124 & 0.0354 & 0.0958 & 0.0357 & -0.0321 & 0.0162 &0.0543 & 0.0659 & 0.0843 & 0.0388 & 0.0026 & -0.0302  \\
\hline
Stat & 0.0180 & 0.0085 & 0.0078 & 0.0039 & 0.0103 & 0.0197 &0.0119 & 0.0067 & 0.0096 & 0.0046 & 0.0077 & 0.0153  \\
\hline
Syst & 0.0102 & 0.0046 & 0.0044 & 0.0022 & 0.0062 & 0.0112 &0.0068 & 0.0038 & 0.0055 & 0.0024 & 0.0041 & 0.0088  \\
\hline
MC Stat & 0.0102 & 0.0044 & 0.0043 & 0.0022 & 0.0062 & 0.0111 &0.0067 & 0.0035 & 0.0054 & 0.0024 & 0.0041 & 0.0086  \\
FSR & 0.0006 & 0.0013 & 0.0004 & 0.0001 & 0.0001 & 0.0007 &0.0006 & 0.0013 & 0.0005 & 0.0002 & - & 0.0008  \\
Eff & 0.0005 & 0.0002 & 0.0001 & - & - & 0.0005 &0.0002 & - & - & - & - & 0.0002  \\
Bkg & 0.0003 & - & 0.0001 & - & - & 0.0003 &0.0001 & - & - & - & - & 0.0001  \\
Smear & - & - & - & - & - & - &- & - & - & - & - & -  \\
PDF & 0.0001 & 0.0003 & 0.0004 & 0.0001 & 0.0004 & 0.0004 &0.0007 & 0.0010 & 0.0011 & 0.0003 & 0.0004 & 0.0017  \\
Extraction & 0.0011 & 0.0002 & 0.0004 & - & 0.0004 & 0.0012 &0.0006 & 0.0001 & 0.0003 & 0.0001 & 0.0003 & 0.0007  \\
\hline
\hline
 & \multicolumn{6}{c}{$y^Z \in [3,3.25]$} & \multicolumn{6}{|c|}{$y^Z \in [3.25,3.6]$}\\
\hline
\hline
Coefficient & $A_{0}$ & $A_{1}$ & $A_{2}$ & $A_{3}$ &$\Delta A_{4}$ & $A_{0}-A{2}$ & $A_{0}$ & $A_{1}$ & $A_{2}$ & $A_{3}$ &$\Delta A_{4}$ & $A_{0}-A{2}$\\
\hline
Total & 0.0708 & 0.0665 & 0.0778 & 0.0144 & 0.0029 & -0.0070 &0.0443 & 0.0723 & 0.0583 & 0.0341 & 0.0171 & -0.0139  \\
\hline
Stat & 0.0107 & 0.0068 & 0.0110 & 0.0051 & 0.0075 & 0.0154 &0.0102 & 0.0065 & 0.0102 & 0.0047 & 0.0072 & 0.0145  \\
\hline
Syst & 0.0058 & 0.0039 & 0.0064 & 0.0028 & 0.0040 & 0.0087 &0.0053 & 0.0036 & 0.0059 & 0.0026 & 0.0037 & 0.0078  \\
\hline
MC Stat & 0.0057 & 0.0036 & 0.0063 & 0.0027 & 0.0040 & 0.0085 &0.0052 & 0.0031 & 0.0057 & 0.0026 & 0.0037 & 0.0077  \\
FSR & 0.0009 & 0.0015 & 0.0007 & 0.0003 & - & 0.0012 &0.0007 & 0.0015 & 0.0006 & 0.0002 & - & 0.0010  \\
Eff & 0.0001 & - & 0.0001 & - & - & 0.0002 &0.0001 & - & - & - & - & 0.0002  \\
Bkg & - & - & 0.0001 & - & - & 0.0002 &- & - & - & - & - & -  \\
Smear & - & - & - & - & - & - &- & - & - & - & - & -  \\
PDF & 0.0005 & 0.0004 & 0.0006 & 0.0006 & 0.0004 & 0.0009 &0.0006 & 0.0008 & 0.0008 & 0.0004 & 0.0003 & 0.0005  \\
Extraction & 0.0002 & 0.0001 & 0.0007 & - & 0.0002 & 0.0007 &0.0004 & 0.0001 & 0.0006 & 0.0001 & 0.0002 & 0.0007  \\

\hline
\hline
\end{tabular}}
\label{tab:sys-y}
\end{table}

\clearpage
\begin{table}[!htbp]
\caption{Summary of measured angular coefficients and regularised uncertainties for $A_{2}$, in intervals of \Z-boson \pt, in the region of \Z-boson $y>2$ and different mass ranges. The total systematic uncertainty is shown with the breakdown into its underlying components. Entries marked with `-' indicate that the uncertainty is
below 0.0001.}
\centering
\resizebox{!}{8.25cm}{
\begin{tabular}{c|c|c|c}
\hline
\hline
 & \multicolumn{3}{c}{$\pt(Z) \in [0,3] [\gev]$} \\
\hline
\hline
 & $M_{\mu\mu} \in [55,75]  \ \rm{GeV/c^{2}}$ & $M_{\mu\mu} \in [75,105]  \ \rm{GeV/c^{2}}$ & $M_{\mu\mu} \in [105,120] \ \rm{GeV/c^{2}}$ \\
\hline
Coefficient & $A_{2}$ & $A_{2}$ & $A_{2}$\\
\hline
Total & 0.1931 & 0.0255 & 0.1429  \\
\hline
Stat & 0.0399 & 0.0132 & 0.1204  \\
\hline
Syst & 0.0235 & 0.0077 & 0.0685  \\
\hline
MC Stat & 0.0230 & 0.0075 & 0.0680  \\
FSR & 0.0007 & - & 0.0008  \\
Eff & 0.0012 & 0.0002 & 0.0022  \\
Bkg & 0.0012 & - & 0.0013  \\
Smear & - & - & -  \\
PDF & 0.0028 & 0.0017 & 0.0079  \\
Extraction & 0.0032 & 0.0002 & 0.0017  \\
\hline
\hline
 & \multicolumn{3}{c}{$\pt(Z) \in [3,6] [\gev]$} \\
\hline
\hline
 & $M_{\mu\mu} \in [55,75]  \ \rm{GeV/c^{2}}$ & $M_{\mu\mu} \in [75,105]  \ \rm{GeV/c^{2}}$ & $M_{\mu\mu} \in [105,120] \ \rm{GeV/c^{2}}$ \\
\hline
Coefficient & $A_{2}$ & $A_{2}$ & $A_{2}$\\
\hline
Total & -0.0230 & -0.0122 & -0.0952  \\
\hline
Stat & 0.0323 & 0.0102 & 0.0883  \\
\hline
Syst & 0.0191 & 0.0062 & 0.0485  \\
\hline
MC Stat & 0.0177 & 0.0060 & 0.0483  \\
FSR & 0.0009 & - & 0.0002  \\
Eff & 0.0007 & - & 0.0013  \\
Bkg & 0.0007 & - & 0.0005  \\
Smear & - & - & -  \\
PDF & 0.0062 & 0.0013 & 0.0044  \\
Extraction & 0.0033 & 0.0006 & 0.0028  \\
\hline
\hline
 & \multicolumn{3}{c}{$\pt(Z) \in [6,12] [\gev]$} \\
\hline
\hline
 & $M_{\mu\mu} \in [55,75]  \ \rm{GeV/c^{2}}$ & $M_{\mu\mu} \in [75,105]  \ \rm{GeV/c^{2}}$ & $M_{\mu\mu} \in [105,120] \ \rm{GeV/c^{2}}$ \\
\hline
Coefficient & $A_{2}$ & $A_{2}$ & $A_{2}$\\
\hline
Total & 0.0530 & 0.0335 & 0.0083  \\
\hline
Stat & 0.0274 & 0.0083 & 0.0692  \\
\hline
Syst & 0.0157 & 0.0048 & 0.0398  \\
\hline
MC Stat & 0.0152 & 0.0046 & 0.0390  \\
FSR & 0.0002 & - & 0.0012  \\
Eff & 0.0006 & - & 0.0017  \\
Bkg & 0.0002 & - & 0.0019  \\
Smear & - & - & -  \\
PDF & 0.0019 & 0.0011 & 0.0072  \\
Extraction & 0.0029 & 0.0005 & 0.0019  \\
\hline
\hline
 & \multicolumn{3}{c}{$\pt(Z) \in [12,20] [\gev]$} \\
\hline
\hline
 & $M_{\mu\mu} \in [55,75]  \ \rm{GeV/c^{2}}$ & $M_{\mu\mu} \in [75,105]  \ \rm{GeV/c^{2}}$ & $M_{\mu\mu} \in [105,120] \ \rm{GeV/c^{2}}$ \\
\hline
Coefficient & $A_{2}$ & $A_{2}$ & $A_{2}$\\
\hline
Total & 0.1278 & 0.0583 & 0.1018  \\
\hline
Stat & 0.0291 & 0.0097 & 0.0760  \\
\hline
Syst & 0.0167 & 0.0052 & 0.0449  \\
\hline
MC Stat & 0.0164 & 0.0049 & 0.0422  \\
FSR & 0.0006 & 0.0002 & 0.0013  \\
Eff & 0.0009 & - & 0.0014  \\
Bkg & 0.0007 & 0.0001 & 0.0005  \\
Smear & - & - & -  \\
PDF & 0.0028 & 0.0010 & 0.0151  \\
Extraction & 0.0013 & 0.0012 & 0.0028  \\
\hline
\hline
\end{tabular}}
\label{tab:sys-a2}
\end{table}


\clearpage



\newpage
\centerline
{\large\bf LHCb collaboration}
\begin
{flushleft}
\small
R.~Aaij$^{32}$,
A.S.W.~Abdelmotteleb$^{56}$,
C.~Abell{\'a}n~Beteta$^{50}$,
F.~Abudin{\'e}n$^{56}$,
T.~Ackernley$^{60}$,
B.~Adeva$^{46}$,
M.~Adinolfi$^{54}$,
H.~Afsharnia$^{9}$,
C.~Agapopoulou$^{13}$,
C.A.~Aidala$^{86}$
% ...

{\footnotesize \it

% ...
$^{9}$Universit{\'e} Clermont Auvergne, CNRS/IN2P3, LPC, Clermont-Ferrand, France\\
% ...
$^{13}$LPNHE, Sorbonne Universit{\'e}, Paris Diderot Sorbonne Paris Cit{\'e}, CNRS/IN2P3, Paris, France\\
% ...
$^{32}$Nikhef National Institute for Subatomic Physics, Amsterdam, Netherlands\\
% ...
$^{46}$Instituto Galego de F{\'\i}sica de Altas Enerx{\'\i}as (IGFAE), Universidade de Santiago de Compostela, Santiago de Compostela, Spain\\
% ...
$^{50}$Physik-Institut, Universit{\"a}t Z{\"u}rich, Z{\"u}rich, Switzerland\\
% ...
$^{54}$H.H. Wills Physics Laboratory, University of Bristol, Bristol, United Kingdom\\
% ...
$^{56}$Department of Physics, University of Warwick, Coventry, United Kingdom\\
% ...
$^{60}$Oliver Lodge Laboratory, University of Liverpool, Liverpool, United Kingdom\\
% ...
$^{68}$Syracuse University, Syracuse, NY, United States\\
% ...
$^{86}$University of Michigan, Ann Arbor, United States, associated to $^{68}$\\
\bigskip
% ...
}
\end{flushleft}

\newpage
\appendix

\end{document}